\documentclass[draftclsnofoot, conference]{IEEEtran}

\usepackage{cite}

\ifCLASSINFOpdf
\else
\fi

\usepackage{array}
\usepackage[table]{xcolor}
\usepackage{multirow}

\ifCLASSOPTIONcompsoc
  \usepackage[caption=false,font=normalsize,labelfont=sf,textfont=sf]{subfig}
\else
  \usepackage[caption=false,font=footnotesize]{subfig}
\fi

\usepackage{xurl}

\usepackage[breaklinks]{hyperref}

\usepackage{subfiles}
\usepackage{tabularx}
\usepackage{graphicx}

\begin{document}
%
\title{Designing Game Feel.\\A Survey}

\author{\IEEEauthorblockN{Martin Pichlmair and Mads Johansen}
\IEEEauthorblockA{Center for Computer Games Research\\
IT University Copenhagen\\
Copenhagen, Denmark\\
Email: mpic@itu.dk and madj@itu.dk}}


%

\maketitle

\begin{abstract}
Game feel design is the intentional design of the affective impact of moment-to-moment interaction with games. In this paper we survey academic research and publications by practitioners to give a complete overview of the state of research concerning this aspect of game design, including context from related areas. We analysed over 200 sources and categorised their content according to the design purpose presented. This resulted in three different domains of intended player experiences: physicality, amplification, and support. In these domains, the act of polishing that determines game feel, takes the shape of tuning, juicing, and streamlining respectively. Tuning the physicality of game objects creates cohesion, predictability, and the resulting movement informs many other design aspects. Juicing is the act of polishing amplification and it results in empowerment and provides clarity of feedback by communicating the importance of game events. Streamlining allows a game to act on the intention of the player, supporting the execution of actions in the game. These three design intents are the main means through which designers control minute details of interactivity and inform the player's reaction. This framework and its nuanced vocabulary can lead to an understanding of game feel that is shared between practitioners and researchers as highlighted in the concluding future research section.
\end{abstract}



\section{Introduction}

The logical starting point for any exploration of game feel as a research subject is the book of the same name by Steve Swink, who defines Game Feel as ``real-time control of virtual objects in a simulated space, with interactions emphasised by polish'' \cite{swinkGameFeel2009}. He further expands on that definition by stating that great-feeling games convey five kinds of experiences, namely:

\begin{itemize}
\item{The aesthetic sensation of control}
\item{The pleasure of learning, practising and mastering a skill}
\item{Extension of the senses}
\item{Extension of identity}
\item{Interaction with a unique physical reality within the game}
\end{itemize}

Yet, while Swink’s definition of game feel covers a wide range of video games, it is too limited to encompass all kinds of them. He explicitly excludes particular classes of games from the group of games that possess the quality of game feel. Doug Wilson \cite{wilsonTaleTwoJousts2016} challenges this aspect of Swink’s book and extends the notion of aesthetic sensation of control by connecting game feel with the cultural history of gestures. Wilson distinguishes between `Game Feel' and `game feel', the first being the positive feeling of control that Swink describes, the second being any feeling a game communicates. Jesse Schell does not mention the term `game feel' in his book \cite{schellArtGameDesign2008} at all. Yet he writes that designers should consider how their game feels in the context of required skills, learnability, and balance. The journalist and game maker Tim Rogers wrote an exhaustive article \cite{rogersPraiseStickyFriction2010} about what he calls `friction', an alternative term for how a game feels. Friction is the experience of the player pushing against the boundaries of the system. It is the feeling of the inertia of the design working against the user’s force. Friction is often experienced by the player over a longer duration than moment-to-moment interaction, as it extends over several game elements. Friction can be the defining element of a game, or just a part of the experience. Rogers’ monolithic article is difficult to parse and, while detailed, does not go deep into specific aspects of game feel. It mostly recounts different feelings the author had during particular situations in games, establishing a wide vocabulary for talking about the aesthetic experience of playing. Similarly, Anthropy \& Clark \cite{anthropyGameDesignVocabulary2014} establish a `game design vocabulary' in their book of the same name. In it, they approach friction from the designer’s perspective, and call it `resistance'. The resistance of the game determines the experienced friction. It decides how the game feels to the player.

Ehrndal \cite{ehrndalHolisticApproachDesigning2012} approaches the topic of game feel in an academic way that links reflections of practitioners with aesthetic theories of games. Larsen \cite{larsenCollisionThrillsUnpacking2016} starts from a similar point and attempts to define an `aesthetics of action'. He builds on Swink \cite{swinkGameFeel2009} and Nijman \cite{nijmanArtScreenshake2013}, both game developers more than researchers, in order to analyse the components of a game that contribute to what he calls a `thrilling experience'. Yang \cite{yangQueeringGameFeel2018} on the other hand expands the theory of game feel to include the metaphorical aspects of game objects and their relations to players. Building on queer theory, he includes political aspects of games in their `feel' in order to communicate the diversity of the gameplay experience over diverse players, and also in order to provide game makers with a richer set of design tools.

While games are multi-sensory experiences, we are focusing on the haptic and visual aspects of game feel in this article, aware that narrative content, sound, music, art, and many other aspects of a game influence how it feels. Very similar techniques to the ones described in this paper exist for example for designing the feel of the story of a game \cite{vonnegutShapesStories1985}, writing the voices of in-game characters, balancing its rules and tuning its atmosphere. Yet, this paper is concerned with moment-to-moment interactivity \cite{kumariRoleUncertaintyMomenttoMoment2019,sivakGAME3400Level2012,swinkGameFeelSecret2007}, microinteractions \cite{safferMicrointeractionsDesigningDetails2013} and interactions with core loops \cite{sicartLoopsMetagamesUnderstanding2015} and their design. Unlike Swink’s precise but narrow definition of `Game Feel', we will look at game feel more broadly as the affective aspect of real-time interactivity.

\section{The Physicality of Interactivity}

This survey paper gives an overview of the history, context, and state of the art of the understanding of game feel and how to design it. It is based on research in the field and publications by practitioners in order to capture both, conceptual and the practical knowledge. This chapter gives an overview of academic lines of thinking that lead to an understanding of game feel.

\subsection{From Flow to Feel}

In the years before Steve Swink wrote the book `Game Feel' in 2009, research about the link between emotions and gameplay was most often connected to Mihaly Csikszentmihalyi's famous `Flow theory', which was one of the results of a global research project about experiences that are ``so gratifying that people are willing to do it for its own sake, with little concern for what they will get out of it, even when it is difficult or dangerous'' \cite{csikszentmihalyiFlowPsychologyOptimal1990}. Tasks that allow for this quality of experience feature the following eight elements:

\begin{enumerate}
	\item a task that can be completed;
	\item the ability to concentrate on the task;
	\item that concentration is possible because the task has clear goals;
	\item that concentration is possible because the task provides immediate feedback;
	\item the ability to exercise a sense of control over actions;
	\item a deep but effortless involvement that removes awareness of the frustrations of everyday life;
	\item concern for self disappears, but sense of self emerges stronger afterwards; and
	\item the sense of the duration of time is altered.
\end{enumerate}

The concept of flow has had an enormous influence on the understanding of experiential qualities of games. Sweetser \& Wyeth \cite{sweetserGameFlowModelEvaluating2005} adapted Flow theory to games and Juul \cite{juulHalfRealVideoGames2005} discusses and criticises the theory's relevance for describing enjoyable challenges in games. Both texts contain multiple references to how games make a player feel, so it is natural to think of it as a stepping stone towards a closer examination of game feel. Ciccoricco \cite{ciccoriccoNarrativeCognitionFlow2012} links Flow to the gameplay experience of Mirror’s Edge \cite{diceMirrorEdge2008}, a game that was sold on its merits in fluidity of movement, and contrasts it with the feminist concept of fluidity. Jenova Chen \cite{chenFlowGamesEverything2007} famously not only based his graduation thesis on Flow but also released three successful commercial games based on his understanding of this concept with his studio That Game Company; Flow \cite{thatgamecompanyFlow2006}, Flower \cite{thatgamecompanyFlower2009}, and Journey \cite{thatgamecompanyJourney2012}.

Game feel is most strongly reliant on points three, four, and five in the above list of criteria. Clarity of goals will be discussed in the context of streamlining of the player experience in Section \ref{sec:Support}. Immediate feedback is at the heart of this paper and of the link between a game and how it feels. The sense of control will be mentioned at various points, for example in relation to the illusion of control and immersion. While some of the other items in Csikszentmihalyi's list apply to games too, they do so in a more indirect way.

\subsection{The Purpose of Juice}

While Flow is very well suited for understanding the dynamics of immersion, game feel is more focused on the role interactivity plays in this process. A design concept often mentioned when talking about how interactivity can be intensified is `juice'. Juice amplifies interactivity by providing excessive amounts of feedback in relation to user input \cite{juulGoodFeedbackBad2016} (see also \cite{hicksGoodGameFeel2018}). The goal of juice is to make actions feel significant. It is superfluous from a strictly mechanical perspective, but turns interacting with the system into a more pleasurable experience. There is an adequacy to juice. Juice-rich interaction makes it hard to learn what aspects of interactivity have mechanical importance \cite{doucetOilItSpoil2016} as decoding the actual system behind the game is cumbersome, when the whole screen is filled with wobbling particle effects --- unless this is a conscious aesthetic choice and itself part of the game’s mechanics. 

At the same time, the diversity of the medium allows that some games --- we can call them `toys' or `autotelic experiences' \cite{sicartPlayMatters2014} --- are almost purely made of juice. Interacting with those toys is still playful and based on feedback amplified by juice. Only through feedback can we learn (to play), and all play is learning (\cite{geeGoodVideoGames2007,kosterTheoryFunGame2005,sutton-smithAmbiguityPlay1997}, see also \cite{iacovidesGamePlayBreakdownsBreakthroughs2015} and \cite{pichlmairDesigningEmotionsArguments2004}). One could even go so far as to argue that all cognition is rooted in feedback from the real world that we actively engage with in a process of interactive cognition (see \cite{gedenrydHowDesignersWork1998}, based on \cite{deweyQuestCertaintyStudy1930}).

Overall, the goal of the application of juiciness is to enhance the feedback when interacting with game objects. Kao \cite{kaoEffectsJuicinessAction2020} conducted a large-scale study on the amount of juice appropriate for a specific gaming situation, concluding that juice has to be applied adequately to the situation. In their study, medium and high levels of juiciness outperformed extreme levels and the absence of juiciness across the measures of player experience, intrinsic motivation, play time, and in-game performance. Hicks et al. \cite{hicksGoodGameFeel2018} bridge industry knowledge and academic analysis, building on Juul’s \cite{juulCasualRevolutionMIT2009,juulGoodFeedbackBad2016}, Schell’s \cite{schellArtGameDesign2008} and Deterding’s \cite{deterdingLensIntrinsicSkill2015} work on juice in video games (see also \cite{atanasovJuicinessExploringDesigning2013}). They present a framework for analysis of juiciness in games that they hope can also be used in game design. 
Sometimes juice exists not for the player but for the audience watching the game. Rogers hints at that when he says ``The player knows where the hit range of the weapon is. He doesn't see the little juice-dance of the chain-daggers.'' \cite{rogersPraiseStickyFriction2010}. Swink attributes a similar effect to ragdoll physics: ``The ragdoll raptors have `over the shoulder' appeal. People walking by someone playing the game often stop and want to know more...'' \cite{swinkGameFeel2009}. Gage \cite{gageBuildingGamesThat2018} describes the upsides of having a game that is readable on different levels in his talk on `subway legibility'. More generally, there are elements of juiciness that designers implement for the audience, especially for streaming and e-Sports \cite{carlssonDesigningSpectatorInterfaces2015}. Some elements might also draw in the player but become invisible to them over time.

Hunicke \cite{hunickeLovingYourPlayer2009} remarks that ``juiciness can be applied to abstract forms and elements and it is a way of embodying arbitrarily defined objects and giving them some aliveness, some qua, some thing, some tenderness''. Interestingly, Swink \cite{swinkGameFeel2009}, Larsen \cite{larsenCollisionThrillsUnpacking2016} as well as Fullerton \cite{fullertonGameDesignWorkshop2014} use the term `polishing' to describe something very similar to this. Fullerton describes the act of polishing as ``the impression of physicality created by layering of reactive motion, proactive motion, sounds, and effects, and the synergy between those layers'' \cite{fullertonGameDesignWorkshop2014}. In other words she sees polish as a means of giving physicality to inanimate objects in order to render them more tangible, which is remarkably similar to Hunicke’s reasoning for juiciness. Practitioners use the term `polish' closer to its dictionary sense. They call many things polish, e.g. fixing the timing of voice cues, or fixing bugs in the code (see e.g. \cite{suddabyImportantWaysAdd2013}). Polish is linked to juiciness in that all juicy elements are polished at some point, but it is seen as a mostly aesthetic endeavour that stops short of changing the basic rules of a game, its core narrative, or its principal game mechanics. In practice, this separation is not always maintained and the connection between juicing, polishing, designing, and the feelings elicited by the feedback loop of interacting with a game is complex.

The intentionality of polishing and juicing apparent in Hunicke’s and Fullerton’s comments is at the centre of Lisa Brown’s assertion that ``you’re not juicing your game –-- you’re actually picking a feeling that your game should communicate and juicing that feeling''  \cite{brownNuanceJuiceTalk2016}. In this paper we limit ourselves to feelings, steering clear of complex emotions --- love, hate, and such –-- which require a closer connection between the game and the player than the moment-to-moment interaction we are concerned with provides. Baumeister et al. \cite{baumeisterHowEmotionShapes2007} call the class of feelings we work with `automatic affect'. This type of affect is closely linked to experience via feedback loops. The emotional reaction to a stimulus has an effect on future experiences of stimuli and those have an effect on the person experiencing them and so on. Affect is generally characterised by arousal, the quality of the experience, and valence, which can be either positive or negative. Game designers are of course concerned with positive as well as negative thoughts, because stretches of sadness and near-frustrating challenges provide the perfect breeding ground for happiness and relief. It is important to note that humans are capable of experiencing multiple and even conflicting emotions simultaneously \cite{larsenCanPeopleFeel2001}. Further, experiments in mood regulation have shown that humans exhibit a `homeostatic mood management mechanism' \cite{forgasManagingMoodsEvidence2002}. After initial mood-congruent responses, we spontaneously reverse and replace those by mood-incongruent reactions. So, additionally to the feedback between the outside world --- including mediated experiences like video games --- and our emotional state, there is a feedback loop built into our mood.

\subsection{Designing for Emotions}

The connection between emotion and cognition is a vast research field and proponents of that field like Okon-Singer et al. often speak of how central emotion is to our cognition \cite{okon-singerNeurobiologyEmotionCognition2015}. The emotional aspect of design has been reflected by design thinkers like Löwgren, Kirkpatrick, Hodent, and Karhulati. Löwgren \cite{lowgrenPliabilityExperientialQuality2007,lowgrenArticulationInteractionEsthetics2009} provides a vocabulary for linking aesthetics, design, and emotional responses. The sensibility and precision he employs to talk about design elements and design choices is valuable for better discussions about game design. Hodent \cite{hodentSkillBuildingSeriesEmotion2020} successfully bridges the chasm between interaction design and user experience design for games and accurately summarises the links between Norman's work \cite{normanEmotionalDesignWhy2004} and video game design. In general, User Experience (UX) Design is an area that is concerned with the experiential aspects of interactivity within the vast field of Human-Computer Interaction. Hassenzahl \cite{hassenzahlNeedsAffectInteractive2010} presents an in-depth study of the complex links between needs, affect, and interactivity. Methods originating in UX design have found their way into games \cite{longWhatGamesUser17}. Their main influence is indirect. They inform the design and iteration process by offering a portfolio of tools and techniques. For example, Dan Saffer’s proposal of microinteractions \cite{safferMicrointeractionsDesigningDetails2013} links to game design in that the basic components of a microinteractions are triggers, rules, feedback, and loops (or modes) --- all basic building blocks of game design.

Kirkpatrick \cite{kirkpatrickArtGamenessCritical2007,kirkpatrickControllerHandScreen2009,kirkpatrickAestheticTheoryVideo2011} and Karhulahti \cite{karhulahtiKinestheticTheoryVideogames2013} employ aesthetic theory and critical theory to build an aesthetic theory of video games that encompasses kinaesthetics as a foundational building block. Keogh \cite{keoghPlayBodiesHow2018} takes this argument a step further in that he argues that the phenomenology of play, rooted in the understanding of embodiment by Merleau-Ponty \cite{merleau-pontyPhenomenologyPerception1982}, Bateson \cite{batesonStepsEcologyMind1972}, and Weiss \cite{weissBodyImagesEmbodiment1999}, ``must not start with the experience of the player’s body, but with the experience through which the player’s amalgam embodiment in and as part of the videogame performance emerges.'' \cite{keoghPlayBodiesHow2018}. Surman \cite{surmanPleasureSpectacleReward2007}, Davnall \cite{davnallbeckyDrJohnsonSore2016}, Putney \cite{putneyPraiseSunYoga2016} offer three personal takes on three different games, echoing similar struggles of coming to terms with the bodily experience of playing games and the implications of the act of doing so.

In summary, game feel research is concerned with how our minds and our bodies experience the emotions of playing games. The question of how to design the emotional aspect of the play experience has been at the centre of a lot of research that connects design theory, psychology, phenomenology, philosophy, and many more areas.

\section{Game Feel Design Elements in Practice}

Gameplay designers, of which some have a programming background and some have a design background, have analysed their own practice in countless blog posts, podcast episodes, conference presentations, and, sometimes, scholarly publications. The majority of works concerned with topics of game feel are descriptive in nature. They usually focus on either a single game or a specific feature or set of features that the designers have worked on. What can be learned from these texts, more than anything, is that experienced gameplay designers are very conscious about which aspects of their game are relevant for shaping its feel.

Some practitioners talk about how to structure game development processes around the design of game feel \cite{grayHowPrototypeGame2005,perrySingleMostUseful2013,brownSecretsGameFeel2015}. Others focus on giving broad overviews of techniques \cite{swinkGameFeelSecret2007,jonassonJuiceItLose2012,nijmanArtScreenshake2013,forestieBestPracticesFast2018,forestieHowDesignFeedback2019}. Podcasts and video series by experienced practitioners such as The Spelunky Showlike \cite{suttnerSpelunkyShowlike2018,suttnerSpelunkyShowlike36,suttnerSpelunkyShowlike38,suttnerSpelunkyShowlike39,suttnerSpelunkyShowlike42}, The Clark Tank \cite{clarkClarkTank2019} or Game Maker’s Toolkit \cite{brownSecretsGameFeel2015,brownWhyDoesCeleste2019}, frequently discuss game feel design as a part of their coverage of game design topics.

There’s a noticeable lack of big picture thinking among practitioners, with a couple of exceptions. Hodent \cite{hodentSkillBuildingSeriesEmotion2020} links game feel to classical concepts of game development like the `3Cs' \cite{mcenteeRaymanOrigins2012}, User Experience Design, as well as to Norman's theories on emotional design \cite{normanEmotionalDesignWhy2004}. Song \cite{songImprovingCombatImpact2005} provides an excellent overview of how to model the feeling of impact in action games. Turner \cite{turnerOhMyThat2015} wrote one of the few articles on how to influence game feel via sound design rooted in his own work in game audio. Ismail \cite{ismailSixStagesGame2015} writes about community development, explaining how communities of makers establish more and more sophisticated discourse about their practice over time. Another text by a practitioner that contextualises game feel in wider political and social development is Yang’s \cite{yangQueeringGameFeel2018} essay about Queering Game Feel.

In general, the topics that these practical articles cover cannot easily be isolated from each other. They all concern feedback and how it relates to controls of a game. If the game is regarded as a feedback system (following \cite{cookWhatAreGame2006,cookLoopsArcs2012} and \cite{hunickeMDAFormalApproach2004}), then game feel can be seen as a modulation of said feedback system. Designing game feel is designing the adequate feedback for eliciting a specific feeling or affective reaction. The following chapters list different design elements that determine the game feel, the feel of moment-to-moment interaction. We cluster design elements into classes according to the game’s subsystem the designer is discussing. Table \ref{tab:overview} presents an overview of the areas we're looking at and lists the most relevant examples mentioned. The table is not an exhaustive overview of all aspects of game feel from a practitioner’s perspective. It is a starting point for going deeper into practices most relevant for designing game feel.

\begin{table}[!tb]
\renewcommand{\arraystretch}{1.25} 
    \caption{Game feel design elements overview.}
    \centering
    \begin{tabular}{l|c|c|c|l}
         \textbf{Design Element} &
         \parbox[t]{3mm}{\rotatebox[origin=c]{90}{\textbf{ Physicality }}} & 
         \parbox[t]{3mm}{\rotatebox[origin=c]{90}{\textbf{ Amplification }}} & 
         \parbox[t]{3mm}{\rotatebox[origin=c]{90}{\textbf{ Support }}} &
         \textbf{Key References} \\
    \hline
    \multicolumn{5}{l}{\textbf{Movement and Actions}} \\
    \hline
        Basic Movement  & \textbullet\ &   &   & \cite{dahlMeasuringHowGame2015,normoyleTradeoffsResponsivenessNaturalness2014, pignolePlatformerControlsHow2014,fasterholdtYouSayJump2016,pittmanMathGameProgrammers2016,coneItRocketScience2018} \\
        Gravity         & \textbullet\ &   &   & \cite{fasterholdtYouSayJump2016, lefkyAccelerationDueGravity2007, alessiGamesDemystifiedSuper2008} \\
        Terminal Velocity          &   &   & \textbullet\ & \cite{fasterholdtYouSayJump2016} \\
        Coyote Time                &   &   & \textbullet\ & \cite{saltsmanTuningCanabalt2010, venturelliGameFeelTips2014} \\
        Invincibility Frames       &   &   & \textbullet\ & \cite{smashpediaInvincibilityFrame, mora-zamoraIntegratedFrameworkGame2019, siuProgrammingModelBoss2016} \\
        Corner Correction          &   &   & \textbullet\ & \cite{gilbertMovementMechanics2012, doucetOilItSpoil2016} \\
        Collision Shapes           & \textbullet\ &   & \textbullet\ & \cite{wiltshireHowHitboxesWork2020} \\
        Button Caching             &   &   & \textbullet\ & \cite{fasterholdtYouSayJump2016} \\
        Spring-locked Modes        & \textbullet\ &   &   & \cite{raskinHumaneInterfaceNew2000, johnsonModesSurveyResults1989, anthropyGameDesignVocabulary2014} \\
        Assisted Aiming           &   &   & \textbullet\ & \cite{doucetOilItSpoil2016, kayaliTwoHalvesPlaySimulation2008,zimmermanReadingPlayerMind2010} \\
    \hline
    \multicolumn{5}{l}{\textbf{Event Signification}} \\
    \hline
        Screen Shake                & \textbullet\ & \textbullet\ & \textbullet\ & \cite{nijmanArtScreenshake2013, jonassonJuiceItLose2012, pennerRobertPennerProgramming2002a,sitnikEasingFunctionsCheat2020} \\
        Knock-back \& Recoil        & \textbullet\ & \textbullet\ &   & \cite{nijmanArtScreenshake2013, barlogWhyKratosAxe2018} \\
        One-shot Particle Effects   &   & \textbullet\ & \textbullet\ & \cite{reevesParticleSystemsTechnique1983,ilmonenSecondOrderParticle2003,lattaBuildingMillionParticleSystem2004, lovatoSqueezingMoreJuice2015, jonassonJuiceItLose2012, rockenbeckInFAMOUSSecondSon2014, vainioVisualEffectsInFAMOUS2014} \\
        Cooldown Visualisation      &   &   & \textbullet\ & \cite{CooldownsCanBe,griesemerDesignNumbersCooldowns2012,kingPrincipalsUIDesign2019,babichUIUXDesign2019} \\
        Ragdoll Physics             & \textbullet\ & \textbullet\ & \textbullet\ & \cite{jakobsenAdvancedCharacterPhysics2001, swinkGameFeel2009} \\
        Colour Flashing             &   & \textbullet\ &  & \cite{perrySingleMostUseful2013, kaoExploringImpactAvatar2016, nijmanArtScreenshake2013} \\
        Impact Markers              &   & \textbullet\ & \textbullet\ & \cite{stephensonUXAnalysisFirstPerson2018, songImprovingCombatImpact2005} \\
        Hit Stop                    &   & \textbullet\ & \textbullet\ & \cite{danielsWhyGamesFeel2007,hurricaneImpactFreeze2010,songImprovingCombatImpact2005, brownWhyDoesCeleste2019} \\
        Audio Feedback              & \textbullet\ & \textbullet\ & \textbullet\ & \cite{berbeceGameFeelWhy2015, franinovicSonicInteractionDesign2013, nackePlayerGameInteractionAffective2011} \\
        Haptic Feedback             & \textbullet\ & \textbullet\ & \textbullet\ & \cite{orozcoRoleHapticsGames2012, songImprovingCombatImpact2005} \\
    
    \hline
    \multicolumn{5}{l}{\textbf{Time Manipulation}} \\
    \hline
        Freeze Frames              &   & \textbullet\ & \textbullet\ & \cite{songImprovingCombatImpact2005} \\
        Slow Motion                &   & \textbullet\ & \textbullet\ & \cite{songImprovingCombatImpact2005} \\
        Bullet Time                &   & \textbullet\ & \textbullet\ & \cite{sabbaghArtDesigningVisceral2015, porterVideogameHistoryBullettime2010} \\
        Instant Replays            &   & \textbullet\ & \textbullet\ & \cite{songImprovingCombatImpact2005} \\
        
    \hline
    \multicolumn{5}{l}{\textbf{Persistence}} \\
    \hline
        Trails                     &   &   & \textbullet\ & \cite{reevesParticleSystemsTechnique1983, berbeceGameFeelWhy2015,nuttMagicTowerFallDepth2015} \\
        Decals \& Debris           &   &   & \textbullet\ & \cite{birdwellCabalValveDesign1999, berbeceGameFeelWhy2015} \\
        Follow-Through             & \textbullet\ &   &   & \cite{thomasIllusionLifeDisney1981} \\
        Fluid Interfaces           & \textbullet\ &   & \textbullet\ & \cite{karunamuniDesigningFluidInterfaces2018, gitterBuildingFluidInterfaces2018, changAnimationCartoonsUser1993} \\
        Idle Animations            &   &   & \textbullet\ & \cite{alexanderQuietImportanceIdle2019,coutureWhatMakesGreat2018,thomasIllusionLifeDisney1981} \\
        
    \hline
    \multicolumn{5}{l}{\textbf{Scene Framing}} \\
    \hline
        Points of Interest        &   &   & \textbullet\ & \cite{kerenScrollBackTheory2015,kerenGamasutraItayKeren2015,meyerITSPCameraExplained2013} \\
        Dynamic Camera            &   & \textbullet\ & \textbullet\ & \cite{christieCameraControlComputer2008,haigh-hutchinsonFundamentalsRealTimeCamera2005,haigh-hutchinsonRealTimeCameras2009,haigh-hutchinsonRealTimeCamerasNavigation2009,burelliVirtualCinematographyGames2013,yannakakisAffectiveCameraControl2010,perrySingleMostUseful2013} \\
    \hline
    \end{tabular}
    \label{tab:overview}
\end{table}

\subsection{Movement and Actions}

\begin{figure}[t]
    \centering
    \includegraphics[width=\linewidth]{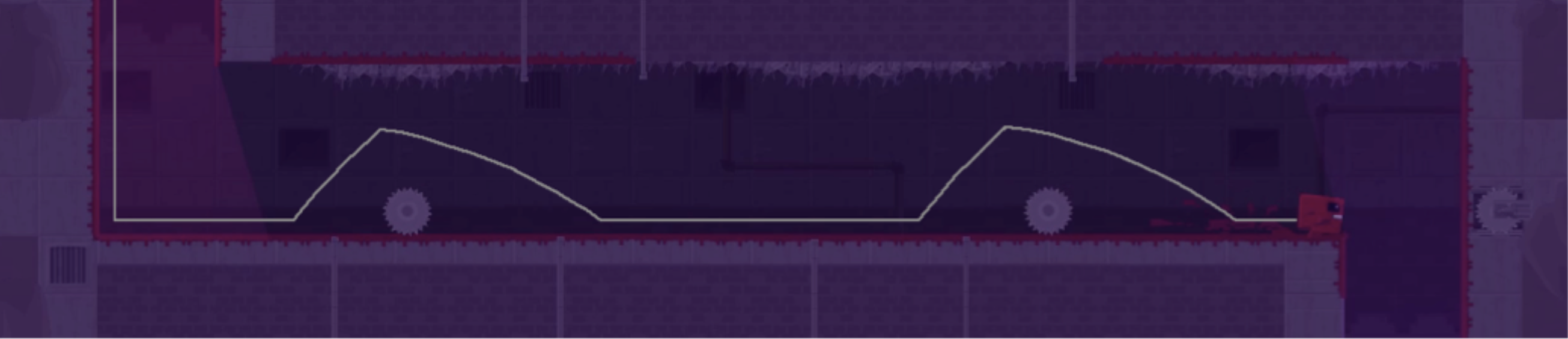}
    \caption{Super Meat Boy allows the player to interrupt a jump, when the jump button is released, to avoid ceiling elements. Image from \cite{fasterholdtYouSayJump2016}}
    \label{fig:jumpcurve}
\end{figure}

The first category of design elements is concerned with movement of the character and other objects and with what happens if the character or an object collide. Controlling an on-screen character means navigating the game world and interacting with other characters and objects. Most writing on this aspect of game feel is concerned with 2D games. Dahl \& Kraus \cite{dahlMeasuringHowGame2015} provide a good starting point for exploring this topic. Normoyle and Jörg \cite{normoyleTradeoffsResponsivenessNaturalness2014} look at the trade-off between naturalness of movement and responsiveness of controls. Pignole \cite{pignolePlatformerControlsHow2014} describes 10 different aspects of how to design controls that feel responsive. While purely grounded in his own experience, these recommendations are easy to pick up and adapt to any game with 2D character movement. In a more extensive study, Fasterhold et al. \cite{fasterholdtYouSayJump2016} provide an overview of parameters for modelling running and jumping in games. This paper also contains an extensive literature review and insights into implementation details of various platformer games. The authors’ model features 21 different parameters to describe basic 2D movement. The key argument in this paper is that movement parameters afford \cite{normanDesignEverydayThings1988} level patterns. Mario’s \cite{nintendoSuperMarioBros1988a} jump curve, for example, excellently facilitates precision descends thanks to featuring terminal velocity that makes future positions easier to predict. Super Meat Boy \cite{teammeatSuperMeatBoy2010}, as another example, abruptly interrupts a jump when the jump button is released, which makes hazardous ceiling elements a viable level design choice, since they can be avoided more easily than if the jump would continue. This is shown in Fig. \ref{fig:jumpcurve}.

Hamaluik \cite{hamaluikSuperMarioWorld2012} used screen recordings to measure and reconstruct all relevant parameters for Super Mario World \cite{hamaluikSuperMarioWorld2012}. Game Makers’ Toolkit (\cite{brownWhyDoesCeleste2019}, see also \cite{thorsonLevelDesignWorkshop2017}) runs a more informal analysis of the platformer Celeste \cite{mattmakesgamesCeleste2018}. Celeste’s player controller’s source code was published \cite{berryCelestePlayerController2018}, allowing for even deeper analysis. Fiedler \cite{fiedlerIntegrationBasics2004} provides good starting points for implementing advanced controls and simulations.

\subsubsection{Basic Movement}

This design element is concerned with the most basic parameters defining the interactive movement of an on-screen object, in most cases the player character. The parameters in question are speed, acceleration, friction, and breaking speed (see \cite{dahlMeasuringHowGame2015}, \cite{normoyleTradeoffsResponsivenessNaturalness2014} and \cite{pignolePlatformerControlsHow2014}). If the player can jump, the strength of the jump force as well as eventual air friction come into play, too. In the case of 2D games, Fasterhold et al. \cite{fasterholdtYouSayJump2016} list these and more parameters and how they are related. Saltsman \cite{saltsmanTuningCanabalt2010} covers movement in one specific platformer, Canabalt \cite{saltsmanCanabalt2009}, in greater detail. Pittman \cite{pittmanMathGameProgrammers2016} explains the mathematics behind jump mechanics. The exact requirements for tuning the movement of a game is often so deeply connected to the gameplay that it is hard to generalise. An in-depth analysis of the car ball game hybrid Rocket League \cite{psyonixRocketLeague2015} is  presented by Cone \cite{coneItRocketScience2018} and demonstrates how steering of a vehicle is tuned in similar ways to platformer movement. 

\subsubsection{Gravity}

The strength of gravity defines how much force pushes an object towards the ground. Games rarely feature earth-like gravity, opting for higher values instead, in order to create a more controlled feeling. Fasterhold et al. \cite{fasterholdtYouSayJump2016} lists the strength of gravity for a number of platformer games. Earth has a gravity of 9.807 m/$s^2$, whereas Super Meat Boy, assuming that the character is 1 meter tall, has a gravity of 41 m/$s^2$ and Super Mario Bros. even features 91.28 m/$s^2$ \cite{lefkyAccelerationDueGravity2007}. Gravity is used as a sophisticated game mechanic in Super Mario Galaxy \cite{nintendoSuperMarioGalaxy2007}, where the character can jump from planet to planet and always aligns appropriately with the surface of the cosmic body. Alessi \cite{alessiGamesDemystifiedSuper2008} wrote up an explanation and prototypical implementation of this gravity system.

\subsubsection{Terminal Velocity}

The existence of terminal velocity in a movement system means that a falling object does not perpetually get faster. It stops to accelerate at a predefined speed, the terminal velocity. As mentioned above, Mario’s \cite{nintendoSuperMarioBros1988a} jump curve \cite{fasterholdtYouSayJump2016} facilitates precision descends thanks to terminal velocity. The additional predictability, that results from the curve becoming a line, supports precision.

\subsubsection{Coyote Time}

\begin{figure}[t]
    \centering
    \includegraphics[width=\linewidth]{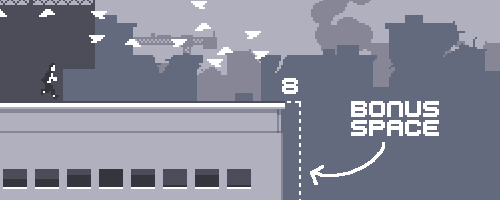}
    \caption{Illustration of `Coyote Time' in Canabalt \cite{saltsmanCanabalt2009}, showing the extra distance from the building where a jump is still accepted. Image from \cite{saltsmanTuningCanabalt2010}}
    \label{fig:coyotetime}
\end{figure}

The term `Coyote Time' refers to a movement system that allows a player to still instigate a jump a short time span after running off a cliff\footnote{The name `Coyote Time' is based on the coyote in the Road Runner series, a character who possessed the power to only fall from a cliff after realising he had been running on thin air for a while.}. It is perhaps the most famous example of supporting the intent of the player. A detailed account of its technical implementation in the minimal platforming game Canabalt \cite{saltsmanCanabalt2009} can be found in \cite{saltsmanTuningCanabalt2010}, shown in Fig. \ref{fig:coyotetime}. Coyote Time is sometimes called `Coyote Jump' or `Ghost Jump', for example in \cite{venturelliGameFeelTips2014}. A similar time-based accessibility feature can be found in Disc Room \cite{calisDiscRoom2020}, where hit boxes that would kill the player get activated only after a delay of up to 50 milliseconds.

\subsubsection{Invincibility Frames}

Short time spans where the player character is invincible. They are a side-effect of player actions like rolling, dodging, respawning, or attacking. SmashPedia \cite{smashpediaInvincibilityFrame} lists 23 different cases of invincibility in Super Smash Bros. Ultimate \cite{nintendoSuperSmashBros2018}, a fighting game, alone. These moments of invincibility are useful for normal players but essential to competitive play and speedrunning. Mora-Zamora and Brenes-Villalobos \cite{mora-zamoraIntegratedFrameworkGame2019} described invincibility frames as a tool for balancing risk and reward. Siu et al. \cite{siuProgrammingModelBoss2016} mention invincibility frames as part of boss fights. The purpose for introducing a few frames of invincibility is usually to support the player, to give them a carefully measured amount of safety that allows them to pull off even more spectacular actions than if they were vulnerable all the time.

\subsubsection{Corner Correction}

Adjusting a character's path if it would otherwise get stuck in level geometry. This is a common convenience in games where walking is a large part of gameplay. Gilbert \cite{gilbertMovementMechanics2012} analyses how it is implemented in The Legend of Zelda \cite{nintendoLegendZelda1986} and Doucet \cite{doucetOilItSpoil2016} offers a detailed analysis of its implementation in Super Mario Bros. 3 \cite{nintendoSuperMarioBros1988a}.

\subsubsection{Collision Shapes}

Collision detection finds overlaps between objects on screen, informing the game when a collision between them occurs. At the heart of collision detection are hit boxes, also called `colliders' with specific shapes\footnote{The authors are aware that this is a slight simplification, but think this description is sufficiently detailed for the purpose of this paper.}. In the case of a 2D game, collision shapes are usually either circles, triangles, or rectangles. In 3D games, they are often spheres, boxes, or capsules. The individual shapes and extents of hit boxes, as well as the coherence between collision shapes and visible game elements, determine how collisions between game elements feel to the player \cite{wiltshireHowHitboxesWork2020}.

\subsubsection{Button Caching}

A common player support function is `Jump Buffering' \cite{fasterholdtYouSayJump2016} (and other forms of button caching), where the controller code buffers the pressing of the jump button for a few frames and executes the jump after the player has landed. Mario \cite{nintendoSuperMarioBros1988} caches the button for 1-2 frames and Braid \cite{numbernoneBraid2008} for 0.23 seconds \cite{fasterholdtYouSayJump2016}.

\subsubsection{Spring-locked Modes}

This is a user interface modality that is actively maintained by the player by pressing and holding a button. The object they are interacting with `switches mode' for the duration that the button is held. This form of interaction is what Raskin \cite{raskinHumaneInterfaceNew2000} calls a `quasimode' and Johnson \& Engelbeck \cite{johnsonModesSurveyResults1989} refer to as `spring-locked mode'. It is often used in order to create anticipation. Games where the player charges an action before unleashing it fall into this category (e.g. Angry Birds \cite{rovioentertainmentAngryBirds2009}, SSX Tricky \cite{bigSSXTricky2001}, R-Type \cite{iremcorporationRType1987}). Exiting the mode can have a specific effect like the charged shot in R-Type (see \cite{alldridgeRTypeIrem19872014}), or it just returns the player to the previous mode, like in the case of Dark Souls \cite{fromsoftwareDarkSouls2009} where the player raises the shield by pushing a button and lowers it by lifting their finger again. Drag and drop is another example of a spring-locked mode that is common in game interfaces.

Further, some games mirror the action of the player and the action of the character. Jumping in the snowboarding game SSX Tricky \cite{bigSSXTricky2001}, for example, is charged by pressing a button and holding it. The character jumps at the moment when the button is released, an aesthetic choice that greatly affects game feel. This implementation creates a relation between the game mechanic and the physical action of the player \cite{anthropyGameDesignVocabulary2014}.

\subsubsection{Assisted Aiming}

Some games help a player with the precision required for aiming. Many shooter games support assisted aiming (e.g. Gears of War \cite{epicgamesGearsWar2006}) and driving games come with countless driving assistance settings\footnote{DiRT 3 \cite{codemastersDiRT2011} features ABS, Dynamic Racing Line, Stability Control, Auto Steer, Corner Braking and Throttle Management --- very similar systems can be found in real cars.}. These features can be regarded as examples of what Doucet calls `oil' \cite{doucetOilItSpoil2016}, the measured exploitation of `illusion of control', as discussed by Kayali \& Purgathofer \cite{kayaliTwoHalvesPlaySimulation2008}. An extensive description of a particular case of assisted aiming for console shooters can be found in Zimmerman \cite{zimmermanReadingPlayerMind2010}.

\subsection{Event Signification}

This class of design elements signifies gameplay-relevant events. Similar techniques are used when events are triggered by the player as when they are triggered by the system. All techniques listed in this section are only active for a limited duration. It is usual to layer several of them depending on the significance and kind of event being communicated.

\subsubsection{Screen Shake}

This effect, which is sometimes also referred to as `camera shake', shakes the camera (or the world) in order to communicate a significant event –-- often an explosion, taking damage, or similar high-impact actions. Nijman \cite{nijmanArtScreenshake2013} and Jonasson \& Purho \cite{jonassonJuiceItLose2012} both mention screen shake. Lerping and easing functions \cite{pennerRobertPennerProgramming2002a,sitnikEasingFunctionsCheat2020} form the technical basis of the implementation of dynamic cinematography like screen shake. Instead of randomly moving the camera, a carefully selected easing function in a semantically significant direction, communicates more information about what has happened, giving the designer more control over what is communicated to the player.

\subsubsection{Recoil}

When the player character is slightly pushed back, e.g. after firing a gun. Nijman \cite{nijmanArtScreenshake2013} describes an implementation in detail, where the firing of a bullet shakes the screen while also pushing the player character a few pixels back, resulting in a side-effect with gameplay implications. A more sophisticated way of achieving something similar is to use inverse kinematics. God of War \cite{readyatdawnstudiosGodWar2018} uses inverse kinematics to model the reaction of the body of the player character when catching his axe \cite{barlogWhyKratosAxe2018}. Not only the arm but the whole body of the character reacts.

\subsubsection{One-shot Particles}

Particle systems \cite{reevesParticleSystemsTechnique1983,ilmonenSecondOrderParticle2003,lattaBuildingMillionParticleSystem2004} are a staple of juicy game design \cite{lovatoSqueezingMoreJuice2015,jonassonJuiceItLose2012}. Practitioners apply them according to context and sophisticated examples feature many layers of particles accompanied by other techniques from this list, like screen shake and sound effects. Some simple particle systems can be faked by using textures (see \cite{nijmanArtScreenshake2013}). Rockenbeck \cite{rockenbeckInFAMOUSSecondSon2014} demonstrates a state-of-the-art particle pipeline and explains how it was used in inFAMOUS: Second Son \cite{suckerpunchproductionsInFAMOUSSecondSon2014}. Vainio \cite{vainioVisualEffectsInFAMOUS2014} describes how this particular system fits into the wider picture of a modern visual effect pipeline.

\subsubsection{Cooldown Visualisation}

Cooldown time is the time it takes after use until and ability in a game can be used again. Its visualisation has to communicate how long the ability is unavailable as well as the moment it becomes available again. Cooldowns are mostly found in role-playing games, where they govern how often spells can be cast or a character ability can be used, and in strategy games where they govern how long it takes to e.g. erect a building. The duration of the cooldown is communicated by greying out the button that triggered an action and gradually revealing it again over the cooldown time. A short overview can be found in \cite{CooldownsCanBe} and \cite{griesemerDesignNumbersCooldowns2012}. A detailed study of optimising the display of cooldowns in a custom user interface can be found in \cite{kingPrincipalsUIDesign2019}. Generally speaking, cooldown visualisations share a lot with progress indicators (see also \cite{babichUIUXDesign2019}).

\subsubsection{Ragdoll Physics}

Modelling a character using joints, forces, and rigid bodies, instead of animations. Switching from animation to ragdoll is a staple for communicating that a character has died. Jakobsen \cite{jakobsenAdvancedCharacterPhysics2001} wrote about this design element before the name `ragdoll' became common. Swink describes \cite{swinkGameFeel2009} how they used ragdoll physics in the game Off-road Velociraptor Safari \cite{flashbangstudiosOffroadVelociraptorSafari2008}\cite{dictionVelociraptorMassacre2012}. He also lists a number of games that derive their whole appeal from ragdoll physics.

\subsubsection{Colour Flashing}

This simple but effective technique communicates state changes by overlaying an on-screen graphical object with a colour. Perry \cite{perrySingleMostUseful2013} mentions several practical examples of how to indicate damage or other state changes by e.g. flashing the object colour or flashing the whole screen. A special case is flashing an on-screen element that was destroyed before it gets removed from the screen, a technique that creates persistence over time, which is discussed in regards to several other design elements further down this paper. Research indicates that specific colour choices carry different semantic meanings \cite{kaoExploringImpactAvatar2016}. Nijman \cite{nijmanArtScreenshake2013} demonstrates flashing the enemy sprite white for a frame or two in a 2D platformer to emphasise a hit.

\subsubsection{Impact Markers}

In the absence of a player sprite, for example in first-person games, other visual elements have to be used to indicate events. In action games, especially in shooters, getting shot at is information of prime importance. Stephenson \cite{stephensonUXAnalysisFirstPerson2018} lists several different techniques for signifying direction, kind, and strength of impact, illustrated by game examples. Song \cite{songImprovingCombatImpact2005} explains a number of different elements, most of them covered in their own sections in this overview, specifically for signifying impact. A blend of the colour flashing mentioned above and impact marking is for example achieved with impact lighting, where a light source gets created on impact that illuminates the characters from the point of impact \cite{songImprovingCombatImpact2005}.

\subsubsection{Hit Stop}

Animations pause for a brief moment on impact. This effect, sometimes also called `Impact Freeze', is a staple in fighting and action games and maybe the best researched phenomenon in the area of impact feedback visualisations \cite{danielsWhyGamesFeel2007,hurricaneImpactFreeze2010,songImprovingCombatImpact2005}. Brown \cite{brownWhyDoesCeleste2019} describes frame freezes and their design purposes in Celeste \cite{mattmakesgamesCeleste2018}. Hit stops are usually introduced in order to communicate feedback about the severity of a hit, but can go further than that. Samurai Gunn \cite{teknopantsSamuraiGunn2013} features a subtle variation of impact freeze when a character lands on a platform, `stunning' it for a few frames depending on the height it dropped from. Kratos' axe in God of War \cite{readyatdawnstudiosGodWar2018} freezes when it hits an enemy \cite{barlogWhyKratosAxe2018}.

\subsubsection{Audio Feedback}

Acoustic channels of communication are a very common way of layering information on top of the graphical representation of a game. Apart from supporting immersion, audio can also communicate events that happen off screen. Berbece \cite{berbeceGameFeelWhy2015} not only highlights the importance of sound effects but also explains how to layer several in order to create an easy to read soundscape. Audio feedback in interaction design for games can be regarded as a specific application of Sonic Interaction Design \cite{franinovicSonicInteractionDesign2013}. Nacke and Grimshaw \cite{nackePlayerGameInteractionAffective2011} present research on affective and aesthetic impact of game sound. Overall, sound design is a huge part of game development and offers a rich set of tools and techniques (see e.g., \cite{deighanSoundDesignVideo19}, \cite{marksCompleteGuideGame2009}) that are relevant in relation to game feel but too general to cover in this paper.

\subsubsection{Haptic Feedback}

Haptic feedback, often called `force feedback' or simply `controller vibration', is a standard functionality of console controllers and built into most mobile phones. It is usually used for emphasis and not as a critical component to interactivity. Orozco et al. \cite{orozcoRoleHapticsGames2012} provide a complete overview of the history and significance of haptic feedback for games. Most platform holders have clear guidelines about when to use haptic feedback, which means that platform-exclusive titles often exploit these features better than multi-platform games (see \cite{songImprovingCombatImpact2005}).

\subsection{Time Manipulation}

While hit boxes and movement are spatial, the other dimension often exploited for game feel is time. Zagal and Mateas \cite{zagalTimeVideoGames2010} give a good overview of game time from an analytical standpoint. The design intent of time manipulation is most often to amplify the experience or to clarify the intensity or direction of an impact. In this section, game time refers to the time of the world simulated in the game whereas real time refers to time in the real world. 

All examples in this game have to do with slowing down or pausing time because games very rarely speed up time. SSX Tricky \cite{bigSSXTricky2001} and Bubble Bobble \cite{taitocorporationBubbleBobble1986} are among the few examples of games that do so. In Drawkanoid \cite{qcfdesignDrawkaniod2018}, a brick destruction game, time speeds up while the player is waiting for the ball to return from a brick’s destruction. No research about speeding up time has been found, so this chapter only covers the rest of the cases of time manipulation.

\subsubsection{Freeze Frames}

The whole screen is frozen for a short duration, often just a few frames. The difference to hit stops, described above, is that those are a localised phenomenon where one or more on-screen objects get paused, excluding them from the temporal flow of the rest of the game, whereas freeze frames technically halt the progression of game time. Song \cite{songImprovingCombatImpact2005} describes how some games pick the best frame of an animation to freeze on and what gameplay implications frame freezes have.

\subsubsection{Slow Motion}

Slowing down game time for a short duration. Whether applied to replays or to linear game time, slow motion helps to communicate events that would otherwise evolve too fast to be fully perceived by the player. A blend of impact freeze and slow motion can be found in Holedown \cite{grapefruktgamesHoledown2018}. The game does not fully freeze on impact, but slows down time to a near halt for a few frames instead. The ability to use slow motion to make an attack look more powerful is mentioned in Song \cite{songImprovingCombatImpact2005}.

\subsubsection{Bullet Time}

Bullet time \cite{sabbaghArtDesigningVisceral2015} is spring-locked slow motion. It serves as a way to pull off more spectacular or precise actions than the player could accomplish in real-time. They empower the player, amplifying their actions. Porter \cite{porterVideogameHistoryBullettime2010} gives an overview of the history of bullet time in movies as well as games. Technically, bullet time is often eased in and out and maintained for a certain amount of time. This can be modelled using attack-decay-sustain-release `ADSR' curves (see \cite{swinkGameFeel2009} for details on their various applications).

Turn-based games like XCOM \cite{firaxisgamesXCOMEnemyUnknown2012} and pause-action games like Fallout \cite{bethesdasoftworksllcFallout2008}, specifically the V.A.T.S. mechanic, pause time while the player queues actions and subsequently advance it in order to show the results of these actions. This pattern of letting a user plan a move without time pressure and then showing a lengthy and potentially intense payoff in real time or even slow motion is quite similar to the temporal dynamics of match-three games like Bejeweled \cite{popcapgamesBejeweled2001}. This particular way of manipulating game time could be regarded as an extreme form of bullet time because it essentially fulfils the same purpose and has the same structure.

\subsubsection{Instant Replay}

Replays of something that has just happened \cite{songImprovingCombatImpact2005}, often slowed down, are triggered automatically. The application of this technique, that originally comes from sports television, is not researched. The fact that replays communicate pivotal moments of gameplay means that replays might help players in identifying moments of importance.

\subsection{Persistence}

Another aspect related to time is persistence, which could also be called `temporal consistency'. Broadly speaking, this cluster of techniques uses spatial representation to communicate time-dependent information. The problem being solved is, in the words of Bay-Wei and Ungar: ``When the user cannot visually track the changes occurring in the interface, the causal connection between the old state of the screen and the new state of the screen is not immediately clear.'' \cite{changAnimationCartoonsUser1993}.

\begin{figure}[t]
    \centering
    \includegraphics[width=\linewidth]{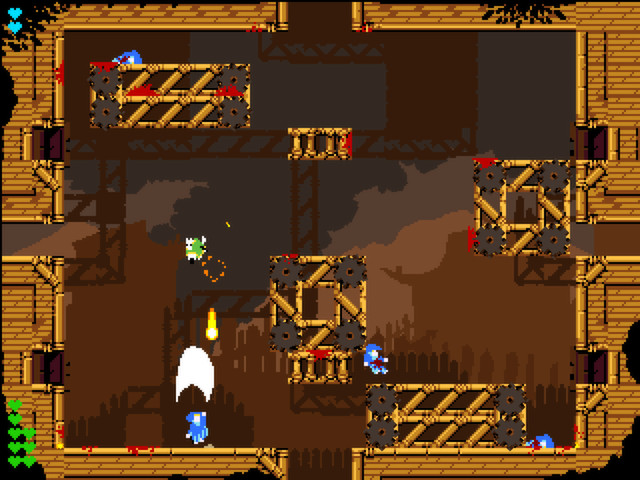}
    \caption{An intense battle in Samurai Gunn \cite{teknopantsSamuraiGunn2013}, the history of motion and battle encoded in the white sword path, the bullet trajectory, smoke particles from where the gun was fired, as well as blood and gore traces all over the level.}
    \label{fig:followThrough}
\end{figure}

From skid marks to particle trails, the purpose of the techniques listed below is always to encode information about the past in the currently displayed image. Even motion blur, which is mentioned further down the list in \ref{sec:cameraTuning}, not only prevents temporal aliasing, but retains the history of movement as a lingering after-image. Very often, the below design elements are used in combination and additionally to other elements that communicate the dynamics of the on-screen action. An example of this can be seen in Fig. \ref{fig:followThrough}.

Temporal consistency also means a consistent frame rate. While this article neither covers technical details nor how bugs and implementation weaknesses affect game feel, it is important to mention that frame rate and especially the duration of the physics time step have a huge influence on how reactive a game feels. Swink also stresses this when he maintains that ``real-time control relies on sustaining three time thresholds: the impression of motion, perceived instantaneous response and continuity of response'' \cite{swinkGameFeel2009}. Fiedler \cite{fiedlerFixYourTimestep2004} provides an excellent introduction into how to implement a stable and reliable feeling core game loop. Cone \cite{coneItRocketScience2018} describes in depth how they solved countless challenges of running a stable physical simulation of the fast moving cars and ball in Rocket League \cite{psyonixRocketLeague2015}.

Overall, temporal consistency techniques are employed in order to allow the player to see either past events or very short events for a longer time. They can be modelled in a diegetic or non-diegetic way (see \cite{iacovidesRemovingHUDImpact2015}), which in this case does not mean that they are modelled in world space or on the interface layer, but rather that the world space is used as an interface layer by attaching trails and particle effects to objects that would be invisible in the real world. The key role of these techniques, from a design perspective, is to support the player.

\subsubsection{Trails}

Traces left behind a moving object. The most prevalent example of temporal persistence in games is found in particle systems \cite{reevesParticleSystemsTechnique1983} and trails. Particle systems that amplify the result of player interaction extend the time that result is visible on screen, creating a dynamic and, for a short while, persistent representation of the player’s interaction history, which could be seen as trails of a player interaction. Particle systems that leave a trail in space as well as time allow the reconstruction of the trajectory of movement of an object. These techniques increase the readability of a scene for the player as well as spectators \cite{berbeceGameFeelWhy2015,nuttMagicTowerFallDepth2015}. 

\subsubsection{Decals \& Debris}

Decals and debris are stationary traces left in the game world. Birdwell \cite{birdwellCabalValveDesign1999} mentions how Valve used decals to acknowledge the actions of the player. Berbece \cite{berbeceGameFeelWhy2015} explains a specific design case, where the player character leaves a blotch of paint after being eliminated from a match, in great detail.

\subsubsection{Follow-Through}

Follow-through, the effect where a part of an animated character or object keeps moving after the main motion has stopped \cite{thomasIllusionLifeDisney1981} is also a way of encoding the history of the motion in subsequent frames. This time, the encoding is not done as a non-diegetic overlay or abstraction, but as movement of parts of the object in question.

\subsubsection{Fluid Interfaces}

Introduced by Apple in 2018 \cite{karunamuniDesigningFluidInterfaces2018}, `Fluid Interfaces' aim at offering more natural interaction forms based on aligning and understanding of intent with physical simulation. They aim at maintaining smooth continuity whenever possible. Gitter \cite{gitterBuildingFluidInterfaces2018} summarises the original presentation and provides a number of code examples. Continuity comes from temporal persistence and spatial coherence, for example when a user interface transition retains aspects of the previous view, while transitioning to a new one \cite{changAnimationCartoonsUser1993}. The same is true for coins that are earned at the end of a round, flying into a virtual purse, each featuring a short trail. Or when representations of pick-ups linger on screen after being collected and then attach to the character.

\subsubsection{Idle Animations}

Small loops of animation that play after a while once the player stops interacting with their character --- when the player character enters the `idle' state \cite{alexanderQuietImportanceIdle2019,coutureWhatMakesGreat2018}. They are superficial in relation to the core mechanics of the game but nevertheless contribute to the overall experience of a game. Idle animations are of course not triggered directly by the player. On the contrary, they are triggered indirectly by not interacting. Idle animations enhance the illusion of life \cite{thomasIllusionLifeDisney1981} of the character.

\subsection{Scene Framing}\label{sec:cameraTuning}

\begin{figure}[t]
    \centering
    \includegraphics[width=\linewidth]{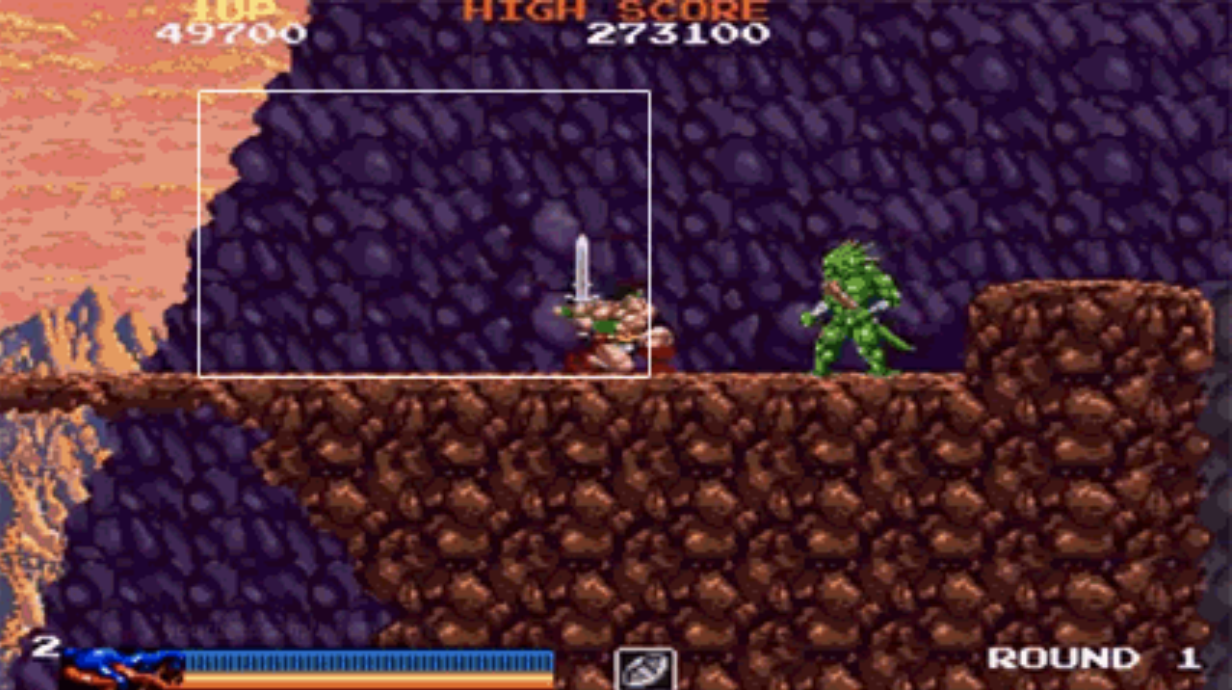}
    \caption{Character camera-window in Rastan Saga \cite{taitocorporationRastanSaga1987}, the camera only moves when Rastan pushes against this window. Image from \cite{kerenGamasutraItayKeren2015}}
    \label{fig:cameraWindow}
    \vspace*{-4pt}
\end{figure}

In racing and flying games, there is a tight link between the field of view, motion blur intensity, and speed. This link defines how the game feels. In 2D games, a variety of techniques are used to enable specific game mechanics, support specific player behaviours, and give specific feedback to players. Keren \cite{kerenScrollBackTheory2015,kerenGamasutraItayKeren2015} assembled a great overview of these techniques, and example of a camera window can be seen in Fig. \ref{fig:cameraWindow}. Eiserloh \cite{eiserlohMathGameProgrammers2016} describes the effects of the maths behind camera controls on game feel from the perspective of a practitioner.

\subsubsection{Points of Interest}

Gameplay-relevant elements highlighted on and off screen. A lot of games have sophisticated techniques to direct the gaze of the player by gradually transitioning the camera focus from the player character to a point of interest. A good example of this is mentioned by Keren \cite{kerenScrollBackTheory2015,kerenGamasutraItayKeren2015} and explained in further detail by Meyer \cite{meyerITSPCameraExplained2013} on hand of his game Insanely Twisted Shadow Planet \cite{shadowplanetproductionsInsanelyTwistedShadow2011}, an exploration game.

\subsubsection{Dynamic Camera}

Articles by Christie et al. \cite{christieCameraControlComputer2008}, Haigh-Hutchinson \cite{haigh-hutchinsonFundamentalsRealTimeCamera2005,haigh-hutchinsonRealTimeCameras2009,haigh-hutchinsonRealTimeCamerasNavigation2009}, and Perry \cite{perrySingleMostUseful2013} provide good starting points for game camera design. Burelli \cite{burelliVirtualCinematographyGames2013} and Yannakakis et al. \cite{yannakakisAffectiveCameraControl2010} examine affective reaction and camera handling. Burelli concludes that interactivity is the key difference between film cinematography and game cinematography, since his study ``demonstrates how the impact on the player experience is mediated by her interaction.“ \cite{burelliVirtualCinematographyGames2013}

\begin{table*}[!t]
\renewcommand{\arraystretch}{1.3}
\caption{Game Feel Design Domains}
\label{tab:domains}
\centering
\begin{tabular}{ |c|c|c|c| }
\hline
\textbf{Design Domain} & Physicality & Amplification & Support \\  
\hline
\textbf{Polishing Task} & Tuning & Juicing & Streamlining \\ 
\hline
\textbf{Description} & 
    \begin{tabular}{@{}c@{}}Setting parameters to specify \\ the behaviour of objects.\end{tabular} &
    \begin{tabular}{@{}c@{}}Adding feedback to \\ emphasise and amplify.\end{tabular} &
    \begin{tabular}{@{}c@{}}Acting on player intent \\ by interpreting the input in \\ context of the gameplay situation.\end{tabular} \\
\hline
\end{tabular}
\end{table*}

\subsection{Summary}

This list of elements of game feel design is by no means exhaustive. We hope that future researchers will use it as a starting point for further exploration of the topic area. Nevertheless, if a game designer concerns themselves with the above design elements and regards them as a collection of methods to draw from, they will be supported in intentionally conveying a game feel they desire.

\section{Game Feel}

The classes of game feel design listed above are connected and, just like Jonasson and Purho keep adding juice in their talk \cite{jonassonJuiceItLose2012}, most moments of interacting with a game are shaped by the presence of several layers of feedback. A good example is the backstabbing attack in Dark Souls \cite{fromsoftwareDarkSouls2009}, an action role-playing game. This attack sequence is triggered by sneaking up to a foe from behind and attacking its back. If successful, the camera locks in place, the enemy and the player character get positioned in predetermined spots relative to each other, and weapon-dependent animations and sounds are played. The player is invincible for the duration of the sequence. The design purpose of this feedback set is to give weight to the effect of a single, but carefully prepared, button press. In general, game designers are most concerned with the quality of interactivity in the core loop \cite{sicartLoopsMetagamesUnderstanding2015} of a game, but that does not mean that they do not employ a lot of the techniques presented in this paper in different parts of the game.

In the following paragraphs we describe three design domains and what the polishing in these domains entails. The domains are physicality, amplification, and support. Polishing means something different for each of them. This diversification of polish helps us to talk about design feel in a precise manner. Table \ref{tab:domains} describes the three domains and their associated polishing task. This is not a complete list of all design aspects of games but intended as a starting point for further research into the relation between design intent and game feel.

\subsection{Tuning Physicality}

The first design domain is the experience of physicality of the system. Swink’s \cite{swinkGameFeel2009} whole concept of Game Feel rests on this pillar. Designers shape the feel of the game by tuning the parameters of the physical simulation \cite{fasterholdtYouSayJump2016}. Depending on how much the game's core loop relies on the joy of movement, attention to detail can become extremely valuable \cite{saltsmanTuningCanabalt2010}. Tuning physicality leads to finely calibrated movement parameters, gravity, and collision shapes. The experience of control is enhanced by additionally applying screen shake, recoil, and knock-back. Appropriate audio design and haptic feedback additionally communicate the physical dynamics of gameplay. It is important to note that for the player experience, it often does not matter whether physicality is simulated or faked. Tweening \cite{burtnykComputerGeneratedKeyFrameAnimation1971}, specifically with easing functions \cite{pennerRobertPennerProgramming2002a,sitnikEasingFunctionsCheat2020}, and various other animation techniques \cite{lasseterPRINCIPLESTRADITIONALANIMATION1987,thomasIllusionLifeDisney1981} can be used to communicate the desired weight of an object. These can be far easier to read as well as implement than a realistic representation. Generally, tuning exploits our knowledge about physicality in order to make interactivity more predictable.

\subsection{Juicing Amplification}

The second design domain is amplification. It primarily serves two purposes: first, it empowers the player. Second, it communicates the importance of events. Empowerment can take many shapes and forms. Bullet time, one of the most iconic ways for amplifying player actions, empowers the player to pull off more precise activities than they could if time progressed linearly. At the same time it also signifies that the player has the opportunity to have greater impact during this time interval than during the rest of the game. Impact freeze on the other hand is mostly used to signify a successful interaction. Charging, which is based on spring-locking, is a technique that balances the reward of empowerment --- the longer the player presses a button, the bigger the impact --- with risk \cite{pichlmairIntentionsExpectationsPlayer2008}.

Juicing the amplification means providing adequate feedback to player actions and creating coherence between different aspects of feedback. Audio, haptic feedback, particle systems, and animation are the most important sources of juiciness. Juice requires exact timing of particle emissions, freeze frames, audio cues, perspective changes, and potentially many more parts of the game. While the power-fantasy aspect of action games thrives on amplification but a lot of playful experiences profit from it. Juicing empowers the player by structuring the reaction of the system to input in a way that amplifies actions adequately.

\subsection{Streamlining Support}\label{sec:Support}

The third design domain is support. It covers techniques that help the player to execute a challenging action or just provide convenience. Doucet \cite{doucetOilItSpoil2016} calls the polishing of support mechanics `oiling', whereas we adopt the less slick term `streamlining' that he also mentions in his article. Streamlining prevents player frustration by making sure that the player receives help where it supports the experience of the game. Doucet lists a couple of examples of how games can be streamlined in order to support the player. The goal of streamlining is to make rough edges of the game disappear, in order to provide a smooth player experience. Most of the time, the player does not want to realise how much the game is supporting them. ``If you do this right then the player wont suspect a thing'' says Pulver \cite{pulverPlatformingLedgeForgiveness2013}. Disc Room's \cite{calisDiscRoom2020} designer Nijman explains that their use of Coyote Time ``has a bunch of good side effects that make it seem like the game knows your intentions.'' \cite{wiltshireHowHitboxesWork2020}. A large portion of the 5400 lines of code that comprises the Celeste character controller is dedicated to providing forgiveness for the player (see \cite{brownWhyDoesCeleste2019} and \cite{thorsonShortThreadFew} for an overview of a few of the features implemented for this purpose). This results in controls that are ``working on the player’s intent rather than making a precise simulation'' \cite{brownWhyDoesCeleste2019}. A much more sophisticated approach was presented by Zimmerman \cite{zimmermanReadingPlayerMind2010} when he describes the selection mechanism for aiming targets when landing on the ground as one-dimensional optimisation problem. 
Elements that enhance temporal consistency offer a different kind of support. Trails that follow projectiles are an example of temporal feedback, since they help to determine the speed and direction of the object by documenting its history. Often particle systems in games have a similar role, and so do skid marks or trails in simulated mud or water. They serve as a visualisation of the past and as an externalisation of information that the player would otherwise have to memorise. That makes them in many cases a service for the player. Continuously displayed game elements like status effects and idle animations similarly take the burden of remembering the state from the player by showing it in the game instead.

Streamlining is not meant to explicitly inform the player about changes in the game’s state. Rather, it is about making the player experience as smooth as adequate. Techniques from User Experience Design \cite{bernhauptEvaluatingUserExperiences2008,isbisterGameUsabilityAdvice2008} can be used for this purpose. A closely related cluster of research concerns game accessibility. While User Experience Design is concerned with setting up game development processes that encompasses user research, the role of accessibility is to widen the audience of games by providing guidelines and tools that make games accessible to players with various kinds of impairments \cite{westinBuildingManifestoGame2015,GameAccessibilityGuidelines}.

\subsection{Designing Game Feel}

Game feel design is minute design work that evokes affect. Affect is the reaction to the concretisation of the expectations towards the feedback of the system. It is subjective and highly dependent on context inside and outside the game. Streamlining, tuning, and even juicing are techniques that help with consciously designing interactive challenges at the heart of the player experience. Game feel is a shortcut for describing how this experience feels. If game feel design is the act of fine-tuning the relationship between expected and actual outcome of an interactive process then it must be regarded central to the game design process.

\section{Future Research}

Game feel is a value-neutral expression. While game designers, as well as scholars, are mostly concerned with what they refer to as `good game feel' (see e.g. \cite{hicksGoodGameFeel2018}), the subjective nature of game feel and the need for ``good negative moments'' \cite{sivakGAME3400Level2012} calls for a more holistic terminology. Those negative moments, if designed consciously, are a valid aesthetic choice, given that ``aesthetics describes the desirable emotional responses evoked in the player, when she interacts with the game system'' \cite{hunickeMDAFormalApproach2004}. Since game feel is the experience of a game’s aesthetic --- following Hunicke’s use of that term --- it spans visual elements, sound design, mechanics, as well as storytelling aspects. Continued exploration of these different game elements and how they contribute to game feel is a worthwhile research endeavour. Sound design stands out as as a field which demands to be researched further. Isolating the aspects that represent polish in different design domains would also be very worthwhile, for example in narrative design.

Tools that help with designing feedback, and game design tools in general, are an area where future research can lead to interesting new experiences. Recent advances have lead to tools for designing feedback \cite{forestieBestPracticesFast2018,forestieHowDesignFeedback2019} or even generating it \cite{johansenSqueezerToolDesigning2020, petterssonSFXR2007}. AI agents and algorithms to help designers analyse and adjust game difficulty \cite{nealenExploringGameSpace2015, isaksenExploringGameSpace2018} or the flow of game play and levels \cite{smithTanagraMixedinitiativeLevel2010, guzdialCoCreativeLevelDesign2018, liapisSentientSketchbookComputerAided2013} have been explored academically. Similar techniques and tools for assisting designers in creating the right feedback are needed.

Research on the effect of feedback on the readability of a game by an autonomous agent is similarly sparse for now. Most of the General Video Game Playing research \cite{levineGeneralVideoGame2013a,khalifaGeneralVideoGame2017,gaina2016TwoPlayerGVGAI2018,perez-liebanaGeneralVideoGame2018, johansenVideoGameDescription2019} is powered by the Video Game Description Language \cite{levineGeneralVideoGame2013a,schaulVideoGameDescription2013}, where feedback is almost non-existent. If feedback can provide support to human players, both in terms of amplification of events and by reducing gameplay rigidity, it might also be able to help AI agents.

Automatic game design tools \cite{cookANGELINAVideogameDesign2017, cookANGELINAVideogameDesign2017a} may also benefit from research into feedback readability and assistance for creating the right feedback. Systems that either partially or fully generate games could benefit from being able to evaluate whether different parts of a game fit together. This applies not just to rules, story, art, sound, etc., but also to feedback. If the feedback of an event is suggesting something different to the player than the rules of the game, then there is a mismatch and the game might feel frustrating or hard to learn (see also \cite{anthropyGameDesignVocabulary2014}).

Designers often work on game feel in very intuitive ways and literature about game feel is very domain-specific. Most written records are by practitioners who are discussing their own projects. And there are large areas that have not been reflected upon. In most cases, these are highly special game mechanics – speeding up time in a game would be an example. Some of the less reflected aspects of game feel-relevant design are of more general relevance, for example the link between audio design and feel. The purpose of this paper is to give an overview of existing research and techniques in order to make this crucial area of game development more accessible to game designers and researchers. That means that less reflected aspects are given less or no weight in this paper. We hope this paper outlines the blind spots and encourages more research into those areas that are neither researched by researchers nor reflected on by practitioners. 

Ultimately, we hope this paper stimulates the creation of more nuanced and reflected design processes, the development of better design tools, and ultimately the design of higher quality interactivity with games. If, as Keogh \cite{keoghIncompleteGameFeel2017} puts it, ``Mechanics are the skeleton. `Polish' or `feel' or `juice' is the meat.'', then a more precise vocabulary is a step towards cooking up better games. Beyond that, the ideas presented in this paper can be applied to a wider range of interactive systems, spreading the sophistication of interaction design in video games to new areas.

\section*{Acknowledgment}

The authors would like to thank Sebastian Risi, Hans-Joachim Backe, Dom Ford, Karin Ryding, S{\'i}lvia Forn{\'o}s, Christian Hviid Mortensen, Charlene Putney, Miruna Vozaru, and Miguel Sicart for their support, feedback, proof reading, and inspiration. We also want to thank Mike Cook, Martin Jonasson, and Steve Swink for great discussions about game feel. And finally the authors want to bow their heads to Petri Purho and Jan Willem Nijman for having great (and ultimately similar) ideas about game feel and continuously talking about them.




\IEEEtriggeratref{199}

\bibliographystyle{IEEEtran}


\end{document}